\documentclass[aps,pra,amssymb,amsfonts,twocolumn]{revtex4}

\newcommand{\ks}{\Phi}
\newcommand{\ksi}[1]{\ks    [#1]}

\begin{document}

  \addtolength{\textheight}{-.3cm}

\title{Counting statistics of interfering Bose-Einstein condensates}

\author{A. L. Shelankov}
\affiliation{Department of Physics, Ume\aa{} University, SE-901 87 Ume\aa{}}
\affiliation{A. F. Ioffe Physico-Technical Institute, 194021 St. Petersburg, Russia}
\author{J. Rammer }
\affiliation{Department of Physics, Ume\aa{} University, SE-901 87 Ume\aa{}}

\begin{abstract}
A method is presented that is able to predict the probability of
outcomes of snapshot measurements, such as the images of the
instantaneous particle density distribution in a quantum many-body
system.  It is shown that a gauge-like transformation of the phase of
the many-body wave function allows one to construct a probability
generating functional, the Fourier transform of which with respect to
the ``gauge'' field returns the joint probability distribution to
detect any given number of particles at various locations.  The method
is applied to the problem of interference of two independent clouds of
Bose-Einstein condensates, where the initially separated clouds with
fixed boson numbers expand and the density profile image of the
overlapping clouds is registered.  In the limit of large particle
numbers, the probability to observe a particular image of the density
profile is shown to be given by a sum of partial probability
distributions, each of which corresponds to a noisy image of
interference of two matter waves with definite phase difference.  In
agreement with earlier theoretical arguments, interference fringes
are, therefore, expected in any single shot measurement, the fringe
pattern randomly varying from run to run.  These results conform to
the physical picture where the Bose-Einstein clouds are in
spontaneously symmetry broken states, the hidden phases of which are
revealed by the density profile measurement via the position of the
interference fringes.  
\end{abstract}


\maketitle

Shortly after the achievement of creating a Bose-Einstein condensate
(BEC) in a dilute gas of ultra-cold atoms, an experiment was performed
where two independent clouds of BEC were brought to overlap in space,
and the particle distribution of atoms measured.  A snapshot picture
of the density of the two overlapping clouds showed interference
fringes analogous to the characteristic interference pattern of a
double-slit experiment \cite{AndTowMie97}.  The text-book double-slit
case is a repeated measurement performed on the same single system,
and the fringes in the arrival pattern of the matter wave emerges only
after many runs.  On the contrary, the former interference fringes are
present in a single snapshot picture of the density distribution of
the overlapping condensates, with a phase, as determined from the
fringe pattern, varying randomly between snapshots.  This experimental
result begs the old question \cite{And86} of how on earth two
independent systems can exhibit knowledge of the relative phase of
their wave functions.  In the present context, this fundamental
problem was analysed in Ref.\cite{JavYoo96} and Ref.\cite{CasDal97}:
Using various models and approaches, indeed it was shown that the
behaviour of condensates with large number of particles is
indistinguishable from the broken symmetry scenario where the
condensate is described by a wave function with a definite, although
unknown, phase.  Since then, the problem has been attracting attention
from many researchers (see
\cite{PetSmi02,PitStr03,Leg06,ImaGriDem07yy}) and references therein).
The purpose of this paper is two-fold: First, we present a method,
which we believe is most suitable for studying density fluctuations in
macroscopic quantum systems, and second, we reanalyse the problem of
interfering BEC clouds using the method.

Interfering BEC clouds provide probably a unique example of a system
with a large number of particles on which a measurement gives
macroscopically distinguishable results that are unpredictable.
Indeed, the interference images registered in identical experiments
differ in the fringe position, and the macroscopic phase extracted
from the image is random from run to run.  For this reason, the
standard description of macroscopic systems, where fluctuations are
small and expectation values are representative, is not sufficient,
and the information we seek is not an expectation value, but the
probabilities for realisations of possible density profiles of BEC
clouds.  Such detailed information is normally not of concern for
macroscopic systems, but is precisely the information needed for
discussing the outcome of a density measurement on interfering BEC's,
viz. to extract a phase from an image, a global feature and an
information which can not be extracted from local expectation values.

Our goal is to obtain directly the statistics of instantaneous
measurements of particle distributions, the kind of information needed
to describe snapshot experiments on many-body systems such as BEC's
\footnote{We take for granted that the optical imaging apparatus used
in the experiments displays the profile of the local density of atoms.
The actual information gathered by a CCD camera is the column density
of the atoms, which can be easily obtained from a 3D profile.}.  Given
a desired spatial resolution $l_{x,y,z}$ in each of the three
dimensions, we imagine a mesh in real space where the elementary
volume is a $l_{x}\times l_{y}\times l_{z}$ box of volume $\Delta
=l_{x} l_{y}l_{z}$.  At the finest level, either a particle or none is
detected in each such pixel. The result of such a position measurement
of the particle locations in a many-body system corresponds thus to
zeros and ones attributed to each such pixel.  Coarse-graining,
\textit{i.e.}, adding the observed particle content of nearby cells,
the extend of which specifies a bin, will attribute a corresponding
{\em integer} number of particles to each bin.  The result of a
snapshot measurement of the positions of the particles is thus
specified by a set of {\em integers},
$\{n\}= n_{1}, n_{2},\ldots,n_{k},\ldots$ the number of
particles detected in the bins located at positions
$\bf{x}_{1}, \bf{x}_{2},\ldots \bf{x}_{k},\ldots$
respectively.  The size of the bin (e.g. the pixel of a CCD camera)
determines the resolution of the image of the particle distribution.

Our aim is to find an efficient method for obtaining the probability
distribution, 
$P(\{n\})$, for the various strings $\{n\}= n_{1},n_{2},\ldots$ of integer
outcomes of the particle distributions for a general quantum system
  consisting of $N$ particles.

 To this end we employ the 
 phase
transformation of the 
wave function 
$\psi({\bf
x}_{1},\ldots,{\bf x}_{N}) \rightarrow \hat{U}_{\lambda}\,\psi({\bf
x}_{1},\ldots,{\bf x}_{N})$ where, similar to a gauge transformation,
\begin{equation}
\hat{U}_{\lambda}\,\psi({\bf x}_{1},\ldots,{\bf x}_{N})
=
e^{i( \lambda ({\bf x}_{1})+\ldots+\lambda ({\bf x}_{N})) }
\psi({\bf x}_{1},\ldots,{\bf x}_{N}) 
.
\label{t0c}
\end{equation}
(A related phase tagging was employed in an investigation of counting
statistics developed for studying electric current noise in mesoscopic
devices \cite{SheRam03}.)  Tagging in this fashion the particles in the
state $| \psi \rangle $, we construct the following functional of the
phase field
\begin{equation}
\ksi{\lambda} = \langle \psi | \hat{U}_{\lambda} | \psi \rangle
.
\label{t1c}
\end{equation}
The Fourier transform of $\ksi{\lambda }$ with respect to the ``gauge
field'' $\lambda (\mathbf{x})$ provides the probability distribution
for strings, 
specifying the possible particle
configurations.  

Indeed, let the real function $\lambda (\mathbf{x})$
be piecewise constant in each of the bins where the number of
particles $n_{k}$ is counted, \textit{i.e.}, 
$\lambda({\bf x})=\lambda_{k}$ in the $k$-th bin. The functional
$\ksi{\lambda}$ becomes a function, $\ks(\{ \lambda \})$ of the variables $\lambda_{1},
\lambda_{2},\ldots \equiv\{\lambda \}$
The probability distribution,
$P(\{n\})$,
is then obtained as the 
 multidimensional
Fourier transform with respect
to the $\lambda$ in each of the bins,
\begin{equation}
P(\{n\}) = \int\limits_{0}^{2\pi} 
{\rm D}\lambda \;
e^{-i  \sum_{k} n_{k} \lambda_{k}}
\ks(\{ \lambda \})
\,,
\label{t3c} 
\end{equation}
where 
${\rm D}\lambda$  denotes the product
 of the differentials,
\begin{equation}
{\rm D}\lambda \equiv\frac{{\rm
     d}\lambda_{1}}{2\pi}\frac{{\rm
     d}\lambda_{2}}{2\pi}\ldots\frac{{\rm d}\lambda_{k}}{2\pi}\ldots
\; ,
\label{chg}
\end{equation}
and the $2\pi $-integration is performed with respect to each of the $\lambda $'s.

To validate this construction, we note that in accordance with
general  rules of quantum mechanics the probability $P(\{n\})
$, \textit{i.e.}, the joint probability that
exactly $n_{k}$ particles are found in the $k$-th
bin, $k=1,2,etc$, is given by the integral,
\begin{equation}
P(\{n\})= 
\int\limits_{\{n\}}{\rm d} \textbf{x}_{1}\ldots {\rm
d}\textbf{x}_{N} |\psi({\bf x}_{1},\ldots,{\bf x}_{N}) |^2\,,
\label{bhg}
\end{equation}
where the region of integration, 
specified by the string $\{n\}= n_{1},
 n_{2},\ldots$ 
is restricted by the set of conditions that for any $k=1,2,etc$ 
exactly $n_{k}$ arguments of the wave function  belong to the $k$-th bin.

The validity of the counting formula Eq.~(\ref{t3c}), where $\ks(\{
\lambda \})$ is given by Eq.~(\ref{t1c} ) with the piecewise constant
$\lambda (\bf{x})$, is immediately seen since the Fourier transform in
Eq.~(\ref{t3c}) has the effect that the unrestricted spatial
integration in Eq.~(\ref{t1c}) is narrowed to the region $\{n\}$ as in
Eq.~(\ref{bhg}).  This concludes our proof that the functional
$\ksi{\lambda}$ is the generating function for the particle
dstributions.

For future needs, we present Eq.~(\ref{t3c}) in a concise form as a
functional integral with respect to the field $\lambda (\textbf{x})$
\begin{equation}
P[n]= \int\limits_{0}^{2\pi }  {\cal D}\lambda \, 
e^{-i\int {\rm d}{\bf x}n({\bf x})\lambda ({\bf x}) } \;
\ksi{\lambda}
\,,
\label{t2c}
\end{equation}
where ${\cal D}\lambda $ is understood as in Eq.~(\ref{chg}), and
$n({\bf x})= \sum_{k} n_k\, \delta ({\bf x} - {\bf x}_{k})$ specifies
an arbitrary assignment of the bins filling or it may be viewed as a
continuous density distribution $n(\bf{x})$.  

We have thus achieved the goal of expressing the statistics of
particle locations, not as customarily directly in terms of the
absolute square of the wave function, but in such a fashion that the
definite outcome of particle content detected in each bin, $n_{k}$,
enters explicitly as parameters.  The spatial resolution of the
profile $n(\mathbf{x})$ is controlled by the smoothness of the field
$\lambda (\mathbf{x})$ one chooses when evaluating the functional
$\lambda $-integration in Eq.~(\ref{t2c}).  The formula,
Eq.~(\ref{t3c}), thus provides the complete counting statistics of
single shot particle distributions for an arbitrary many-body system.
As an application of the method, we consider the counting statistics
of interfering Bose-Einstein condensates. In this example, we restrict
ourselves to the limit of non-interacting particles, neglecting
correlation effects related to the depletion of the condensates.

First, we consider an isolated BEC cloud consisting of $N$ bosons.
For our purposes it will be convenient to build the Fock state of the
cloud in the coherent state representation.  Let $\phi({\bf x})$
denote the state into which condensation occurs, and consider the
coherent state $|\phi_{\theta} \rangle$,
\begin{equation}
|\phi_{\theta}  \rangle = 
e^{
e^{i\theta}
\int {\rm d}{\bf x}\, \phi ({\bf x}) \hat{\psi }^{\dagger}({\bf x}) 
} |0\rangle   
\;,\; 
\int {\rm d}{\bf x}\, |\phi ({\bf x})|^2 = N
\label{75f}
\end{equation}
obtained by operating with the boson field $\hat{\psi }^{\dagger}({\bf
x}) $ on the vacuum state $|0\rangle $.\footnote{ As implied in
Eq.~(\ref{75f}), we choose for calculational convenience the
normalisation of the single-particle wave function to equal the number
of condensed particles, $N$.}  The Fock state $|N \rangle $ where $N$
bosons occupy the single particle mode $\phi( {\bf x})$ is then
obtained by projecting the coherent state, Eq.~(\ref{75f}),
\begin{equation}
|N \rangle = 
c_N
 \int\limits_{0}^{2\pi } \frac{{\rm d} \theta }{2\pi }
\; e^{-i\theta N }
 |\phi_{\theta}  \rangle
\label{fdg}
\end{equation}
where $c_{N}= \sqrt{N!/N^{N}}$ is the normalisation constant.

We then consider two initially separated clouds of bosons with
particle numbers $N_{1}$ and $N_{2}$, and where the particles occupy
single particle states $\phi_{1}({\bf x})$ and $\phi_{2}({\bf x})$,
respectively.  The many-body wave function of the two spatially
non-overlapping clouds is simply the product of the wave functions of
the isolated clouds, and the corresponding quantum state,
$|\Psi\rangle = |N_{1}\rangle_{1}|N_{2}\rangle_{2}$, is thus obtained
expressing the Fock state in terms of the coherent state, according to
\begin{equation}
|\Psi \rangle
=
c_{12}
\int\limits_{0}^{2\pi}\frac{{\rm d} \theta_{1}{\rm d} \theta_{2}}{(2\pi)^2}  
e^{-i (\theta_{1}N_{1}+ \theta_{2}N_{2})}
 e^{
\int {\rm d}{\bf x}\, \psi_{\theta_{1}\theta_{2}}(\mathbf{x})
\hat{\psi }^{\dagger}({\bf x}) 
}
 |0\rangle   
\label{t6f}
\end{equation}
where 
\begin{equation}
\psi_{\theta_{1}\theta_{2}}(\mathbf{x})\equiv
e^{i \theta_{1}}\phi_{1}({\bf x}) 
+
e^{i \theta_{2}}\phi_{2}({\bf x})
\label{jfg}
\end{equation}
and $c_{12}$ is the normalisation constant, $c_{12}=c_{N_1}c_{N_2}$.

One may get the wrong impression that in the quantum state $\Psi $,
Eq.~(\ref{t6f}), the bosons are in the superposition state
$\psi_{\theta_{1}\theta_{2}}(\mathbf{x})$ of Eq.~(\ref{jfg}).
However, unlike genuine superpositions, where a phase transformation
like $\phi_{1} \rightarrow e^{i \varphi }\phi_{1}$ non-trivially
changes the state, the same transformation of the many-body state in
Eq.~(\ref{t6f}) amounts to an overall phase factor multiplication
which does not change the value of any observable.  In fact, the atoms
in the two non-overlapping clouds are not yet even aware that they are
identical species.

In the interference experiment, time evolution provides by the time of
measurement, $t_{0}$, overlap of the initially separated clouds due to
their expansion.  If the atom-atom interaction is neglected, evolution
of the many-body state $|\Psi(t)\rangle$ amounts to the unitary
evolution of the single particle states. Consequently, the expression
Eq.~(\ref{t6f}) is valid for the state $|\Psi(t)\rangle$ provided
$\phi_{1,2}(\mathbf{x})$ are substituted by the evolved
single-particle wave functions $\phi_{1,2}(\mathbf{x},t)$. The initial
states $\phi_{1} $ and $\phi_{2}$ are spatially separated,
\textit{i.e.}, locally the states satisfy
$\phi_{1}(\mathbf{x})\phi_{2}^{*}(\mathbf{x})=0$, and are therefore
trivially orthogonal. It follows from the unitarity of evolution that
the spatially overlapping states $\phi_{1}(\mathbf{x},t)$ and
$\phi_{2}(\mathbf{x},t)$ remain orthogonal at any time.

To simplify notation, we use Eq.~(\ref{t6f}) for the state
$|\Psi(t_{0})\rangle$ of the clouds at the moment of measurement,
$t_{0}$, with the understanding that in all formulae below
$\phi_{1,2}(\mathbf{x})$ stand actually for
$\phi_{1,2}(\mathbf{x},t_{0})$; the functions $\phi_{1}$ and
$\phi_{2}$ are orthogonal and normalised to the particle numbers
$N_{1}$ and $N_{2}$ of the respective clouds.

Having specified the state of the system, we are in the position to
obtain the probability for any possible particle configuration $[n]$
from Eq.~(\ref{t2c}). To evaluate the functional $\ksi{\lambda} =
\langle \Psi | \hat{U}_{\lambda} | \Psi \rangle$ with $|\Psi\rangle$
given by Eq.~(\ref{t6f}), we note that in terms of the field operator,
the tagging of the particles by the phase field via the operation
$\hat{U}_{\lambda }|\Psi\rangle$ is enforced by its transformation
according to $$\hat{\psi}({\bf x}) \rightarrow e^{i \lambda ({\bf x})
} \hat{\psi}({\bf x}).$$ Using the Baker-Hausdorff formula
\cite{ScuZub97} to calculate the matrix element $\langle \Psi |
\hat{U}_{\lambda} | \Psi \rangle$ we arrive after straightforward
calculations at the expression for the particle probability
distribution
\begin{equation}
\begin{array}{lcl}
P[n ({\bf x})]   & =  &    
\rule[-5ex]{0ex}{0ex}
\frac{N_{1}! N_{2}!}{N_{1}^{N_{1}} N_{2}^{N_{2}}}
\int\limits_{0}^{2\pi N}\frac{
{\rm d} \theta {\rm d} \theta'}{(2\pi N)^2} 
\int {\cal D}\lambda  
 \\
  &
 &   
\hspace*{-12ex}\times
\exp\left[\int {\rm d}{\bf x}\,
\left(
e^{i \lambda ({\bf x})}
\psi_{\theta '}^{*}({\bf x}) \psi_{\theta }({\bf x})
-i n({\bf x})\lambda ({\bf x}) 
\right)\right]
\end{array}
\label{i8f}
\end{equation}
where for brevity continuum notation is used, and
\begin{equation}
\psi_{\theta}({\bf x})=
e^{i \nu_{2} \theta }\phi_{1}({\bf x}) 
+
e^{-i \nu_{1} \theta }\phi_{2}({\bf x}) 
\label{v6f}
\end{equation}
with $\nu_{1,2}= N_{1,2}/N$, $N= N_{1}+N_{2}.$\footnote{ An $N$-fold
enlargement of the original integration region of the phases is
convenient to avoid technical problems with the functions $e^{i
\nu_{1,2} \theta }$ which are not $2\pi $-periodic.}  Eq.~(\ref{i8f})
refers only to configurations $n( \mathbf{x})$ which are physical,
\textit{i.e.}, \mbox{ $\int {\rm d} \mathbf{x}\, n(\mathbf{x}) = N $},
and otherwise the probability is identically zero as observed in the
process of obtaining Eq.~(\ref{i8f}) from Eq.~(\ref{t2c}).  For
brevity, in Eq.~(\ref{i8f}) and below we omit the Kronecker symbol
expressing this property.

The formula Eq.~(\ref{i8f}), allows one to find the particle
probability distribution for two Fock state clouds with arbitrary
particle numbers $N_{1}$ and $N_{2}$.

The functional $\lambda $-integral is understood as the
multidimensional one in Eq.~(\ref{t3c}). In our approximation, where
the interaction is ignored, the integral factorises and can be easily
evaluated.  Appropriately choosing the size of the bins, any spatial
resolution for the density profile $n(\mathbf{x})$ can be considered.

For illustration, we first look at the simplest case where each
``cloud'' consists of a single boson, $N_{1}=N_{2}=1$, in the state
$\phi_{1}(\mathbf{x})$ and $\phi_{2}(\mathbf{x})$, respectively.  In
this case, only two types of particle configurations exist: (i) both
particles are located in the bin around point ${\bf x}_{a}$; $\{n({\bf
x})\}_{aa}= 0_{1},0_{2},\ldots,2_{a},\ldots$ where all particle
numbers $n_{k}$ are zero except for bin $a$ where $n_{a}=2$.  (ii) One
of the particles is located in the bin near ${\bf x}_{a}$ and the
other one is in a different bin ${\bf x}_{b}$; $\{n({\bf x})\}_{ab}=
0_{1},0_{2}, \ldots,1_{a},\ldots 1_{b}\ldots$, \textit{i.e.},
$n_{k}=0$ excepting $n_{a}=n_{b}=1$.  Denote by $P_{aa}$ and $P_{ab}$
the probabilities for the corresponding particle configurations, and
in order to single out the dependence on the bin volume present them
as $P_{aa} \equiv {\cal P}_{aa} \Delta^2$ and $P_{ab} \equiv {\cal
P}_{ab} \Delta^2$, assuming the bins small so that
$\phi_{1,2}(\textbf{x})$ do not vary within a bin.  To obtain the
probabilities for these configurations, one integrates in
Eq.~(\ref{i8f}) with respect to the field $\lambda (\mathbf{x})$
assuming that the latter is piecewise constant in each of the bins.
After further integration with respect to the phases $\theta $ and
$\theta '$, one obtains
\begin{eqnarray}
{\cal P}_{aa} &=& 2 
|\phi_{1}(\mathbf{x_{a}})
\phi_{2}(\mathbf{x_{a}})|^2 \, ,
              \label{ifg}\\   
 {\cal P}_{ab} &=& 
 |
 \phi_{1}(\mathbf{x_{a}})\phi_{2}(\mathbf{x_{b}})
 +
 \phi_{1}(\mathbf{x_{b}})\phi_{2}(\mathbf{x_{a}})
 |^2
.
 \label{ifg2}   
\end{eqnarray}
Here, the coefficient of 2 in the expression for ${\cal P}_{aa}$
represents the well-known boson bunching effect \cite{ScuZub97}. The
cross term, $2\Re
\phi_{1}(\mathbf{x_{a}})\phi_{1}^{*}(\mathbf{x_{b}})\phi_{2}(\mathbf{x_{b}})
\phi_{2}^{*}(\mathbf{x_{a}}) $, in ${\cal P}_{ab}$ describes the
expected Hanbury Brown and Twiss effect \cite{ScuZub97}.  Also, one
observes that ${\cal P}_{ab}|_{b \rightarrow a} = 2 {\cal P}_{aa}$;
this is a necessary property to guarantee the correct dependence of
$P_{aa}$ on the choice of the mesh size.

Next we turn to the case of interest, two interfering macroscopic
Bose-Einstein condensates where the number of particles in each cloud
is large, $N_{1,2}\gg 1$.  The main contribution to the phase integral
in Eq.~(\ref{i8f}) comes in that case from the region $\theta '
\approx \theta$, and the particle probability distribution becomes
\begin{equation}
P[n]
= 
\int\limits_{0}^{2\pi }
\frac{{\rm d} \theta}{2\pi } \;
{\cal P}_{\theta}[n]
\label{v08f}
\end{equation}
where
\begin{equation}
\begin{array}{rr}
{\cal P}_{\theta}[n]
= 
\sqrt{2\pi N}
\int {\cal D} \lambda \;
e^{\int {\rm d}{\bf x}\,
\left(
\left(e^{i \lambda ({\bf x})}-1 \right)
\rho_{\theta}({\bf x}) 
-i n({\bf x})\lambda ({\bf x}) 
\right)
}
 \end{array}
\label{x08f}
\end{equation}
with 
 \begin{equation}
 \rho_{\theta  }({\bf x}) =
 \left|
 e^{i \theta }\phi_{1}({\bf x}) + \phi_{2}({\bf x})
  \right|^2 \; .
 \label{x8f}
 \end{equation}

Within the approximation of non-interacting bosons the formula
Eq.~(\ref{v08f}) provides the complete counting statistics of snapshot
particle distribution measurements of two macroscopic interfering
Bose-Einstein condensates in Fock states, \textit{i.e.}, a state where
the particle numbers $N_{1,2}$ of the condensates are large and
separately fixed and their phases are completely indefinite.

To get insight into the content of formula Eq.~(\ref{v08f}), we note
that the distribution $P[n]$ is built by a family of partial
(normalised) distributions ${\cal P}_{\theta }[n]$.  Obviously, the
part of the configuration $[n]$-space where the probability $P[n]$ is
appreciable, is spanned by the corresponding subspaces generated by
the ${\cal P}_{\theta }[n]$'s, $0< \theta < 2\pi $, and we first
discuss the partial contributions ${\cal P}_{\theta }[n]$.  One can
check by comparison that ${\cal P}_{\theta }[n]$ coincides with the
distribution of the outcomes of $N$ independent identical
double-slit-type experiments where the position $\mathbf{x}$ of a
single particle in the superposition state $ \psi_{\theta
}(\mathbf{x})= e^{i \theta }\phi_{1}(\mathbf{x})+\phi_{2}(\mathbf{x})$
is measured.  At the individual pixel level, the distribution of the
number of particles $n_{k}$ registered in the $k$'th bin is Poissonian
reflecting the intrinsic randomness at the level of quantum mechanics.
When typical counts $n_{k}$'s are large, the shot noise fluctuations
in the $n_{k}$'s are relatively small.  Asymptotically, at $N
\rightarrow \infty $, ${\cal P}_{\theta }[n(\mathbf{x})]$ thus selects
the configurations where the density $n(\mathbf{x})$ equals (up to
shot noise) the density expectation value with respect to the wave
function building $\rho_{\theta }(\mathbf{x})$ in Eq.~(\ref{x8f}).
Consequently, the distribution $P[n]$ becomes concentrated on the
(closed) line corresponding to $[n]=\rho_{\theta }$ with $\theta $
running from $0$ to $2\pi $.  Excepting exponentially rare events, the
outcome of any experiment will correspond to a point in the close
vicinity of this line, and consequently, an experiment will always
show the interference picture as if the particles are condensed in a
state with a certain definite phase difference $\theta $
\cite{JavYoo96,CasDal97}.

It is well understood \cite{JavYoo96,CasDal97,PetSmi02,PitStr03,Leg06}
that each experimental run results in a fringe image which corresponds
to a well defined phase difference $\theta $, but its value is random
when the experiment is repeated under identical conditions.  Having at
hand the full counting statistics of the interference image, we are
able to go further into such an analysis and discuss the probability,
$p(\theta )$, of the occurrence of the images corresponding to the
phase $\theta$.  Unlike previous authors, we do not consider it {\em a
priori} obvious that any value of $\theta $ is equally probable as we
are not aware of any physical symmetry that demands states with
different phase difference $\theta$ to be equivalent.  On the
contrary, we observe that the entropy of the partial distribution
${\cal P}_{\theta }$ is $\theta $-dependent,\footnote{ The
distribution ${\cal P}_{\theta }[n]$, Eq.~(\ref{x08f}), is generated
by the product of independent Poissonian counts in the infinitesimal
bins.  The total entropy, $S_{\theta }=- \sum_{[n]}{\cal
P}_{\theta}[n]\ln {\cal P}_{\theta }$, found as the sum with respect
to the bins, equals $S_{\theta } = - \int d {\bf x}\rho_{\theta}({\bf
x})\ln \rho_{\theta}({\bf x})+ const$. } \textit{i.e.}, the
statistical weight of configurations to which a certain value of
$\theta $ can be assigned varies with $\theta $. Besides, there is no
unique and unequivocal algorithm of the assignment, and the question
warrants to be addressed. In general terms, this is a standard problem
of mathematical statistics where one determines the parameters of a
distribution from a sample data \cite{CovTho91}.

Specified a procedure $\Theta[n]$ that assigns the phase difference
$\theta_{[n]}= \Theta [n]$ to a configuration $[n]$, \footnote{For
instance, one may use the assignment procedure where $\theta_{[n]}$ is
evaluated by minimising with respect to $\theta $ the Euclidean
distance, $\int {\rm d} \mathbf{x}\, (n(\mathbf{x})-\rho_{\theta
}(\mathbf{x}))^2$, from a given input distribution $n(\mathbf{x})$ to
$\rho_{\theta }(\mathbf{x})$.} and given the occurrence frequency of
the configurations $P[n]$ in Eq.~(\ref{v08f}), the probability
distribution $p(\theta )$ is found as
\begin{equation}
p(\theta ) = \int\limits_{0}^{2\pi}\frac{{\rm d}\theta'}{2\pi}W_{\theta' }(\theta)
\label{dfg}
\end{equation}
where 
\begin{equation}
W_{\theta' }(\theta) 
= \sum\limits_{[n]}\delta(\theta -\theta_{[n]}){\cal P}_{\theta '}[n] 
\label{gfg}
\end{equation}
is a function of $\theta $, normalised to unity, and with $\theta'$ as
parameter.  For an
unbiased estimator $\Theta[n]$ \cite{CovTho91}, for which
\begin{equation}
\sum_{[n]}\Theta[n]{\cal P}_{\theta}[n] = \theta\, ,
\label{mfg}
\end{equation}
the function $W$ is strongly peaked at $|\theta - \theta '| \sim
\sigma_{\theta '}$, where the width of the peak, $\sigma_{\theta '}$,
can be estimated from the Cram\`{e}r-Rao bound \cite{CovTho91},
\begin{equation}
\sigma_{\theta}^{-2} = \int {\rm d}\mathbf{x}\,
\rho_{\theta}(\mathbf{x}) 
\left(\frac{\partial \ln\rho_{\theta}(\mathbf{x}) }{\partial \theta } \right)^2
\, .
\label{efg}
\end{equation}
The physical meaning of the variance $\sigma_{\theta}$ is that of the
confidence (``the error bars'') with which one is able to estimate the
phase difference from a noisy image.  By virtue of the normalization,
$\rho_{\theta }$ is proportional to the number of particles in the
clouds so that the variance $\sigma_{\theta}$ becomes small in the
macroscopic limit.  Note that the variance varies with $\theta $ in
agreement with the above discussion of the dependence of the
statistical weight on the phase.

In the large-$N$ limit, the function $W_{\theta '}(\theta )$ is a
Gaussian, $$ W_{\theta '}(\theta ) \approx \frac{1}{\sqrt{2\pi
\sigma_{\theta '}^2}} e^{-(\theta -\theta ')^2/ 2\sigma_{\theta
'}^2}\, , $$ the normalisation property with respect to $\theta $
follows from Eq.~(\ref{gfg}).  Although the variance $\sigma_{\theta
'}$ is a function of the phase difference $\theta '$, in the integral
with respect to $\theta '$, Eq.~(\ref{dfg}), one can safely replace
$\sigma_{\theta '} \rightarrow \sigma_{\theta},$ after which the
integral becomes a constant, and $p(\theta )= 1/ 2\pi $ is independent
of $\theta $.  Thus, we see that the statistical inequivalence of the
partial distributions ${\cal P}_{\theta }$ is compatible with the
statement that an image of two overlapping condensates can be
characterised by a phase difference $\theta $, the value of which in
repeated experiments under identical conditions is equally probable in
the interval $0< \theta < 2\pi $.

 Our approach allows us to evaluate the precision $\delta
\theta $ with which the phase difference may be meaningfully assigned
to a noisy fringe image: Using Eq.~(\ref{x8f}), the width of the
``error bars'' given by $\sigma_{\theta }$ in Eq.~(\ref{efg}), can be
expressed via the wave functions $\phi_{1,2}(\bf[x)$ of the
interfering clouds:
\begin{equation}
\sigma_{\theta}^{-2} =
\frac{1}{2}\int {\rm d} \bf{x}\,
\frac{
|\phi_{1}|^2|\phi_{2}|^2 - \Re ( (e^{i \theta }\phi_{1}\phi_{2}^{*})^2)
}{|\phi_{1}|^2 + |\phi_{2}|^2 +
2\Re (e^{i \theta } \phi_{1}\phi_{2}^{*})
}\;.
\label{dhg}
\end{equation}
For a simple illustration, we consider clouds which are initially in
the ground states of 1D harmonic traps centered at $x=\pm \Delta $.
After a free expansion during time $\tau $, the wave functions of the
clouds are $ \phi_{1,2}(x) = \sqrt{N_{1,2}/(a_{\tau }\sqrt{\pi }) }
\exp\left[- (x\pm \Delta )^{2}/(2aa_{\tau }) \right] $ where $a_{\tau
}= a + i \hbar \tau /( ma)$, $m$ and $a$ being the particle mass and
initial width of the clouds, respectively.  If the clouds overlap is
small, \textit{i.e.}, $|a_{\tau }|< \Delta^2/a$ the uncertainty in the
phase $\delta \theta $ is estimated as $\delta
\theta\sim\sigma_{\theta }\sim \exp\left[\Delta ^2/(2a|a_{\tau }|)
\right]/(N_{1}N_{2})^{1/4} $. As no surprise, the uncertainty of the
phase parameter
$\delta \theta \propto (N_{1}N_{2})^{-1/4} $ is small provided $N_{1,2}$ is large enough. 

 In this paper, we have presented a method which allows one to
evaluate the probability of a local density configuration, $[n]=
n(\mathbf{x})$, for a general many-body quantum state.  The
configurations are specified by a string of integers $\{n_{k}\}$,
$k=1,2,etc$ , where the number at the $k$-th position is the number of
particles located in the cell centred at position $\mathbf{x}_{k}$.
The cells cover the relevant part of space, and the size of the cells
defines the resolution of the density distribution. The probability of
a density profile is expressed via the functional integral over the
phase field $\lambda (\mathbf{x})$, Eq.~(\ref{t2c}) or
Eq.~(\ref{t3c}), and the functional $\ksi{\lambda (\mathbf{x})}$
defined by Eqs.~(\ref{t0c}) and (\ref{t1c}).  As usual in the
probability generating function approach, the $n$-point correlation
function can be obtained by taking the functional derivatives of the
functional $\ksi{\lambda (\mathbf{x})}$ with respect to $\lambda
(\mathbf{x}) \rightarrow 0$ at the corresponding $n$ values of the
co-ordinate $\bf x$. We have considered the case of only one species
of particles, the generalisation to many species, bosons or fermions,
is straightforward.

As an example, we have applied the method to the description of the
interference experiment with ultra-cold atoms, where the quantum state
is formed by two overlapping but initially separated clouds of
Bose-Einstein condensates.  We describe the condensates in the
approximation of non-interacting bosons and, in particular, we ignore
the condensate depletion.  
At the level of
self-consistent description, the Gross-Pitaevskii theory, the
interaction only modifies the evolution of the interfering waves
$\phi_{1,2}(\mathbf{x})$ as discussed in \cite{Par08}.  
Beyond the self-consistent approach, the most important effect
of interaction is nontrivial spatial correlations in the noise of
interfering BEC images (see Ref.\cite{ImaGriDem07yy} and references therein).

The density profile of the interfering matter waves generated by the
two independent sources shows strong fluctuations, and the standard
description of macroscopic systems in terms of the expectation value
of the density and its low order, e.g. 2-point, correlators becomes
insufficient.  The ``hidden'' variable of interest is the parameter
$\theta $ which corresponds to the phase difference of the matter
waves in the broken symmetry BEC-state, and this parameter can be
extracted by analysing the density profile, \textit{i.e.}, the CCD
image registered in a snapshot experiment rather than from the
expectation value found by averaging many such measurements.  We have
evaluated the probability for arbitrary density profile realisations
as well as the probability distribution $p(\theta )$ for the phase
difference corresponding to these profiles.  Supporting expectations
of previous authors \cite{JavYoo96,CasDal97,PetSmi02,PitStr03,Leg06},
we have found that the phase difference $\theta $ is distributed
uniformly for an unbiased procedure of assignment of $\theta $ to the
observed images.

In conclusion, we believe that the approach developed in this paper
offers an efficient tool for a detailed analysis of the problem of
interfering matter waves and gives new insight into this fundamental
physical  problem.

\acknowledgments
One of us (J.R.) acknowledges financial support from the Swedish
Research Council.

\end{document}